\documentstyle[12pt]{article}

\font\boldgreek=cmmib10
\mathchardef\mysigma="091B

\def\ben{\begin{enumerate}}  \def\een{\end{enumerate}}
\def\beq{\begin{equation}}   \def\eeq{\end{equation}}
\def\bea{\begin{eqnarray}}  \def\eea{\end{eqnarray}}
\def\nn{\nonumber}
\def\noi{\noindent}

\def\lsim{\raise0.3ex\hbox{$<$\kern-0.75em\raise-1.1ex\hbox{$\sim$}}}
\def\gsim{\raise0.3ex\hbox{$>$\kern-0.75em\raise-1.1ex\hbox{$\sim$}}}

\begin{document}

\begin{center}
{\Large \bf On P-wave meson decay constants}\\
{\Large \bf in the heavy quark limit of QCD}
\par

\vspace{1 truecm}

{\bf A. Le Yaouanc, L. Oliver, O. P\`ene and J.-C. Raynal}\\
{\it Laboratoire de Physique Th\'eorique}\footnote{Unit\'e Mixte de Recherche
UMR 8627 - CNRS }\\    {\it Universit\'e de Paris XI, B\^atiment 210, 91405
Orsay Cedex, France}
\par \vskip 5 truemm
{\bf V. Mor\'enas} \\
{\it Laboratoire de Physique Corpusculaire}\\
{\it Universit\'e Blaise Pascal - CNRS/IN2P3, F-63000 Aubi\`ere Cedex, France}
   \end{center}

\vspace{0.3 truecm}
\begin{abstract}
In previous work it has been shown that, either from a sum rule for 
the subleading Isgur-Wise function $\xi_3(1)$ or
from a combination of Uraltsev and Bjorken SR, one infers for 
$P$-wave states $|\tau_{1/2}(1)| \ll |\tau_{3/2}(1)|$. This
implies, in the heavy quark limit of QCD, a hierarchy for the {\it 
production} rates of $P$-states $\Gamma(\bar{B}_d
\to D \left ( {1 \over 2} \right ) \ell \nu ) \ll \Gamma(\bar{B}_d 
\to D \left ( {3 \over 2} \right ) \ell \nu )$ that
seems at present to be contradicted by experiment. It was also shown 
that the decay constants of $j = {3 \over 2}$
$P$-states vanish in the heavy quark limit of QCD, $f_{3/2}^{(n)} = 
0$. Assuming the {\it model} of factorization in the
decays $\bar{B}_d \to \bar{D}_s^{**}D$, one expects the opposite 
hierarchy for the {\it emission} rates $\Gamma(\bar{B}_d
\to \bar{D}_s \left ( {3 \over 2} \right ) D) \ll \Gamma(\bar{B}_d 
\to \bar{D}_s \left ( {1 \over 2} \right ) D)$, since
$j = {1 \over 2}$ $P$-states are coupled to vacuum. Moreover, using 
Bjorken SR and previously discovered SR involving
heavy-light meson decay constants and IW functions, one can prove that the sums
$\sum\limits_n \left ( {f^{(n)} \over f^{(0)}}\right )^2$, 
$\sum\limits_n \left ( {f_{1/2}^{(n)} \over f^{(0)}}\right )^2$
(where $f^{(n)}$ and $f_{1/2}^{(n)}$ are the decay constants of 
$S$-states and $j = {1\over 2}$ $P$-states) are
divergent. This situation seems to be realized in the relativistic 
quark models \`a la Bakamjian and Thomas, that satisfy
HQET and predict decays constants $f^{(n)}$ and $f_{1/2}^{(n)}$ that 
do not decrease with the radial quantum number $n$.
\end{abstract}

\vspace{0.5 truecm}

\noi LPT Orsay 01-36 \par
\noi June 2001
 
\newpage
\pagestyle{plain}

In previous work \cite{5bisr}, \cite{12r} we have pointed out a 
problem between experiment and the heavy quark limit of
QCD for the semi-leptonic decays of $B$ mesons to $L = 1$ excited 
states $D_{0,1}\left ({1 \over 2}\right )$,
$D_{1,2}\left ( {3 \over 2}\right )$. In a few words, the argument is 
as follows.

 From Bjorken SR \cite{8r} \cite{new4r}

\beq
\label{6e}
		\rho^2 = {1 \over 4}   +  \sum_n 
|\tau^{(n)}_{1/2}(1) |^2 + 2 \sum_n
|\tau^{(n)}_{3/2}(1) |^2 			 \eeq

\noi and the recently demonstrated Uraltsev SR \cite{9r}

  \beq
\label{7e}
		\sum_n  |\tau^{(n)}_{3/2}(1) |^2 - \sum_n 
|\tau^{(n)}_{1/2}(1) |^2 = {1 \over 4}
\quad . \eeq

\noi one obtains,

\bea
\label{8e}
&&		\sum_n  |\tau^{(n)}_{3/2}(1) |^2 = {\rho^2 \over 3} \\
&&		\sum_n  |\tau^{(n)}_{1/2}(1) |^2 = {1 \over 3} \left 
( \rho^2 - {3 \over 4}
\right ) 	\quad .
\label{9e}
\eea

One can see that $\sum\limits_n  |\tau^{(n)}_{3/2}(1) |^2$ is
proportional to $\rho^2$ and that $\sum\limits_n  |\tau^{(n)}_{1/2}(1) |^2$ is
proportional to the {\it deviation} of $\rho^2$ from the lower bound 
${3 \over 4}$.
Then, there is little room left for $\sum\limits_n 
|\tau^{(n)}_{1/2}(1) |^2$, as it has
been pointed out recently from a SR obtained for the subleading 
function $\xi_3(1)$
\cite{8r}. The expected hierarchy for the form factors 
$|\tau_{3/2}(1)| \gg \tau_{1/2}(1)$, that can be inferred
from the precedent equations, implies that $\bar{B}_d \to D^{**}\ell 
\nu$ and $\bar{B}_d \to D^{**}\pi$ (assuming
factorization, a reasonable assumption in view of the recent papers 
on non-leptonic decays with emission of a light
meson) are dominated by the narrow resonances

  \bea
\label{12e}
&& \Gamma \left ( \bar{B}_d \to D_{1,2}\left ( \scriptstyle{3 \over 
2}\right )\ell \nu \right ) \gg \Gamma \left (
\bar{B}_d \to D_{0,1}\left ( \scriptstyle{1 \over 2}\right )\ell \nu \right )\\
&& \Gamma \left ( \bar{B}_d \to
D_{1,2}\left ( \scriptstyle{3 \over 2}\right )\pi \right ) \gg \Gamma 
\left ( \bar{B}_d \to D_{0,1}\left ( \scriptstyle{1
\over 2}\right )\pi \right ) \quad .
\label{number6}
\eea

On the other hand, it was demonstrated that the decay constants of $j 
= {3 \over 2}$ $P$-wave mesons
vanish, $f_{3/2}^{(n)} = 0$, in the heavy quark limit of QCD 
\cite{1r}, \cite{11r}. This result can be summarized in the
statement that we expect the ``broad'' $D^{**}$ resonances ($j = {1 
\over 2}$) to have a much larger decay constant than
the ``narrow'' ones ($j = {3 \over 2}$). This to be contrasted to the 
opposite expected hierarchy for form factors,
$|\tau_{3/2}| \gg |\tau_{1/2}|$ \cite{5bisr}, that can be inferred 
from equations (\ref{8e}), (\ref{9e}). This hierarchy
in the {\it production} is expected to be opposite to the one due to 
the selection rule $f_{3/2}^{(n)} = 0$,

\beq
\label{13e}
\Gamma \left ( \bar{B}_d \to \bar{D}_{s\ 1,2}\left ( \scriptstyle{3 
\over 2}\right )D \right ) \ll \Gamma \left ( \bar{B}_d \to
\bar{D}_{s\ 0,1}\left ( \scriptstyle{1 \over 2}\right ) D \right ) \eeq

\noi where there is {\it emission} of $\bar{D}_{sJ}(j)$. Of course, 
this is only a qualitative statement, because
factorization in the decays (\ref{13e}) is just a model and is not in 
a firm status as in the light meson emission case
(\ref{number6}) \cite{newref}. To summarize, ``broad'' resonances ($j 
= {1 \over 2}$) are expected to be suppressed in the
production, while ``narrow'' resonances ($j = {3 \over 2}$) are 
expected to be suppressed in the emission. The BaBar
experiment has started looking at $B_d \to (\bar{D}K) D$ \cite{13r} 
where such decays with emitted excited states might
be seen, but the statistics has to be improved. \par

Let us now make some further remarks on decay constants of excited 
states. Imposing duality to $\Delta \Gamma$ in the
$B^0_s$-$\bar{B}^0_s$ system in the heavy $b$, $c$ quark limit, the 
following sum rules have been
demonstrated, in the heavy quark limit of QCD \cite{1r}~:

\bea
\label{1e}
		 	&&\sum_n  \ {f^{(n)} \over f^{(0)}}  \ 
\xi^{(n)}(w) = 1				\\
			&&\sum_n  \ {f^{(n)}_{1/2} \over f^{(0)}} \ \tau^{(n)}_{1/2}(w)  = {1 \over 2}
\label{2e}
\eea

\noi The decay constants for $S$-states $f^{(n)}$ and for $j = {1 
\over 2}$ $P$-states
$f^{(n)}_{1/2}$ are properly defined in ref. \cite{1r}. The sum rules
(\ref{1e})-(\ref{2e}) are strong constraints, as they hold for any 
value of $w$. \par

The $w$-dependence in eqns. (\ref{1e}), (\ref{2e}) was obtained by 
considering the two-body intermediate states of the
type $D_s\bar{D}_s$ (ground state and excited states) for the 
diagrams of the spectator quark type. As explained in ref.
\cite{1r}, in this kind of diagrams the transition $\bar{B}_s \to 
D_s\bar{D}_s \to B_s$ occurs by $\bar{D}_s$ emission and
$D_s$ production by the $\bar{B}_s$ ($\bar{s}$ quark being 
spectator), followed by $D_s$ emission and $\bar{D}_s$
production by the $B_s$. In the heavy quark limit, the expression for 
the contribution of one intermediate state is
proportional to a quantity of the type $f_{D_s}^2[\xi (w_c)]^2$ where 
$f_{D_s}$ is a generic ground state or excited $D_s$
meson decay constant and $\xi (w_c)$ is a generic Isgur-Wise function 
$\xi^{(n)}(w)$, $\tau_{1/2}^{(n)}(w)$ or
$\tau_{3/2}^{(n)}(w)$ \cite{new4r}, taken at the fixed value of $q^2 = m_c^2$~:

\beq
\label{3e}
w_c = {m_b^2 + m_c^2 - q^2 \over 2 m_b \ m_c} = {m_b \over 2 m_c} \quad .
\eeq

\noi {\it Varying} the ratio $m_b/m_c$, the $w$-dependence in the IW 
functions appears. Identifying the sum over the
exclusive modes with the contribution to the quark box diagram having 
the same topology \cite{3r}, one obtains,
considering the vector or the axial weak current, the sum rules 
(\ref{1e}) and (\ref{2e}). \par

In an earlier paper \cite{4r}, we had demonstrated that indeed 
duality for $\Delta \Gamma_{B_s}$ occurs for $N_c = 3$ in
the heavy quark limit in the particular Shifman-Voloshin conditions 
$\Lambda_{QCD} \ll m_b - m_c \ll m_b, m_c$
\cite{5r}. \par

On the other hand, the sum rules (\ref{1e}), (\ref{2e}) were studied 
\cite{6r} within the relativistic quark models of
the Bakamjian-Thomas (BT) type \cite{10r}, that satisfy HQET 
relations for form factors and decay constants. The sum rules
are satisfied for different values of $w$, although the numerical 
convergence becomes worst as $w$ increases.

 From (\ref{1e}), (\ref{2e}), using Schwartz inequality, one can 
obtain the lower bounds~:

\bea
\label{4e}
&&	\left [\sum_n \left [ \xi^{(n)}(w)\right ]^2 \right ] \left [ \sum_n
\left ( {f^{(n)} \over f^{(0)}} \right )^2 \right ] \geq  \left  ( 
\sum_n {f^{(n)}
\over f^{(0)}} \xi^{(n)}(w) \right )^2  = 1	 \\
&&\left [ \sum_n \left [ \tau^{(n)}_{1/2}(w) \right ]^2 \right ]\left [ \sum_n
\left ( {f^{(n)}_{1/2} \over f^{(0)}} \right )^2 \right ] \geq 
\left  ( \sum_n
{f^{(n)}_{1/2} \over f^{(0)}}   \tau^{(n)}_{1/2}(w) \right )^2  = {1 \over 4}
\label{5e}
\eea

\noi Considering first these inequalities at $w = 1$, one can see, from
$\xi^{(n)}(1)  = 0$ for $n \not= 0$, that (\ref{4e}) does not provide 
any useful constraint. However, from (\ref{5e}) at
$w = 1$ and (\ref{9e}) one obtains the bound for the $j={1 \over 2}$ 
$P$-wave decay constants

\beq
\sum_n \left ({f^{(n)}_{1/2} \over f^{(0)}} \right )^2  \geq {3 \over 
4\rho^2-3}
  	 \label{10e}
\eeq

\noi that contains the IW slope $\rho^2$ in the right-hand side. 
Although the bound (\ref{10e}) is very weak, it deserves
a few comments. Its r.h.s., and also its l.h.s. diverge as $\rho^2 
\to {3 \over 4}$ and it reflects the fact that the
excited $P$-wave mesons $D_{0,1}(j = {1 \over 2})$ do indeed couple 
to vacuum, respectively through the vector and axial
currents for $J = 0, 1$, as was already proved from the sum rule 
(\ref{2e}). This is to be contrasted with the selection
rule $f_{3/2}^{(n)} = 0$ \cite{1r} \cite{11r} that applies to the 
excited mesons $D_{1,2}(j = {3 \over 2})$. In the
example of the non-relativistic quark model, also given as an 
illustration in ref. \cite{1r}, both sides of the inequality
(\ref{10e}) are of $O(v^2/c^2)$, since $f_{1/2}^{(n)}$ is of $O(v/c)$ 
(formula (17) of \cite{1r}) and $\rho^2$ is of
$O(c^2/v^2)$ (formula (52) of \cite{1r}). In this case, as can be 
easily seen using completeness, the l.h.s. of
(\ref{10e}) is infinite, proportional to the divergent integral 
$\int_0^{\infty} p^4 dp$, two powers of $p$ coming from
the current and $p^2$ from the measure d$\vec{p}$. \par

Actually, one can demonstrate this divergence also in field theory, 
in the heavy quark limit of QCD by considering
arbitrary large $w$. From Bjorken SR for any value of $w$ \cite{new4r},

\beq
\label{number1}
{w+1 \over 2} \sum_n |\xi^{(n)}(w) |^2 + (w-1) \left [ 2  \sum_n |
\tau_{1/2}^{(n)}(w)|^2 + (w+1)^2  \sum_n |\tau_{3/2}^{(n)}(w)|^2 
\right ] + \cdots = 1
\eeq

\noi one obtains, since all terms in this sum are definite positive

\beq
\label{number2}
\sum_n |\xi^{(n)}(w) |^2 \leq {2  \over w+1} \qquad  \sum_n |
\tau_{1/2}^{(n)}(w)|^2 \leq {1 \over 2(w-1)} \
\eeq

\noi From (\ref{4e}) and (\ref{5e}) for any $w$ one gets

\beq
\label{number3}
\sum_n \left ( {f^{(n)} \over f^{(0)}}\right )^2 \geq {w+1 \over 2} 
\qquad  \sum_n \left (
{f_{1/2}^{(n)} \over f^{(0)}}\right )^2 \geq {w-1 \over 2}
\eeq

\noi Since the l.h.s. of these inequalities is independent of $w$, 
that can be made arbitrarily large, one concludes
that the sums $\sum\limits_n \left ( {f^{(n)} \over f^{(0)}}\right 
)^2$, $\sum\limits_n \left (
{f_{1/2}^{(n)} \over f^{(0)}}\right )^2$ must diverge. \par

This situation seems to be realized in quark models \`a la
Bakamjian and Thomas. The decay constants $f^{(n)}$ and 
$f^{(n)}_{1/2}$ were computed within the BT quark
models for different quark-antiquark potentials up to $n = 10$, and 
the convergence of
the SR (\ref{1e}) and (\ref{2e}) was studied for different values of 
$w$ \cite{6r}.
The error on the decay constants induced by a truncation procedure in 
the calculation
increases strongly with $n$. The decay constants $f^{(n)}$ and 
$f^{(n)}_{1/2}$ do not
decrease with increasing $n$. For the most sophisticated 
Godfrey-Isgur potential, that describes all $q\bar{q}$,
$q\bar{Q}$ and $Q\bar{Q}$ quarkonia, including spin dependent effects 
\cite{14r}, one obtains the decay constants of the
Table for the $S$-states and the $j = {1 \over 2}$ $P$-states \cite{6r}.

\newpage
\begin{center}
\begin{tabular}{|c|c|c|}
\hline
\hline
Radial excitation 	&$\sqrt{M} f^{(n)}$ (GeV$^{3/2}$) &$\sqrt{M} 
f_{1/2}^{(n)}$ (GeV$^{3/2}$)  \\
\hline
$n = 0$	&0.67(2) &0.64(2) \\
$n = 1$	&0.73(4) &0.66(4) \\
$n = 2$ &0.76(5) &0.71(5) \\
$n = 3$ &0.78(9) &0.73(8)\\
$n = 4$ &0.80(10) &0.76(11)\\
\hline
\hline
  \end{tabular}
\end{center}

{\baselineskip = 12 pt
\noindent {\bf Table : } Decay constants $f^{(n)}$, $f_{1/2}^{(n)}$ 
for radial excitations $S$-states and $j = {1 \over
2}$ $P$-states in the GI spectroscopic model. The estimated error is 
in parentheses \protect{\cite{6r}}, and $M$ is the
bound state mass.}\par \vskip 5 truemm

In view of the values of the decay constants in the Table, besides 
the qualitative hierarchy (\ref{13e}), one expects
for the $n = 0$ states, assuming factorization, neglecting Penguin 
diagrams and also the mass differences between the $P$
states and the ground state~:

\bea
\label{number4}
&&{\Gamma(\bar{B}_d \to D_{s0}^- \left ( {1 \over 2} \right ) D^+) 
\over \Gamma(\bar{B}_d \to D_s^- D^+ )}
\cong {\Gamma(\bar{B}_d \to D_{s0}^- \left ( {1 \over 2} \right ) 
D^{*+}) \over \Gamma(\bar{B}_d \to
D_s^- D^{*+} )} \cong 1 \nn \\
&&{\Gamma(\bar{B}_d \to D_{s1}^- \left ( {1 \over 2} \right ) D^+) 
\over \Gamma(\bar{B}_d \to D_s^{*-} D^+ )}
\cong {\Gamma(\bar{B}_d \to D_{s1}^- \left ( {1 \over 2} \right ) 
D^{*+}) \over \Gamma(\bar{B}_d \to
D_s^{*-} D^{*+} )} \cong 1 \ .
\eea

\noi These relations follow from the approximate numerical equality 
$f^{(0)} \cong f_{1/2}^{(0)}$ in the Table and the
heavy quark relations among, respectively, the decay constants of 
$D$, $D^*$ and $D_{0}\left ( {1 \over 2}\right )$,
$D_{1}\left ( {1 \over 2}\right )$ mesons \cite{1r}. \par

In conclusion, we have underlined a disymmetry in the prediction of 
the rates of production and emission of $P$-wave
heavy-light mesons, and we have undertaken a theoretical discussion 
of decay constants of excited heavy mesons.


\newpage
\begin{thebibliography}{99}

\bibitem{5bisr} A. Le Yaouanc, D. Melikhov, L. Oliver, O. P\`ene, V. 
Mor\'enas and J.-C. Raynal, Phys. Lett. {\bf B480},
119 (2000).

\bibitem{12r} A. Le Yaouanc, L. Oliver, O. P\`ene, V. Mor\'enas and 
J.-C. Raynal, LPT Orsay 01-20 (March 2001).

\bibitem{8r} J. D. Bjorken, invited talk at Les Rencontres de 
Physique de la Vall\'ee d'Aoste, La Thuile (SLAC Report
No. SLAC-PUB-5278, 1990 (unpublished)~; J. D. Bjorken, I.
Dunietz and J. Taron, Nucl. Phys. {\bf B371}, 111 (1992).

\bibitem{new4r} N. Isgur and M. B. Wise, Phys. Rev. {\bf D43}, 819 (1991).

\bibitem{9r} N. Uraltsev, Phys. Lett. {\bf B501}, 86 (2001).

\bibitem{1r} A. Le Yaouanc, L. Oliver, O. P\`ene and J.-C. Raynal, 
Phys. Lett. {\bf B387}, 582 (1996).

\bibitem{11r} S. Veseli and I. Dunietz, Phys. Rev. {\bf D54}, 6803 (1996).

\bibitem{newref} M. Beneke, G. Buchalla, M. Neubert and C. T. 
Sachrajda, Nucl. Phys. {\bf B591}, 313 (2000).

\bibitem{13r} Sophie Trincaz, Th\`ese de Doctorat de l'Universit\'e 
de Paris VI, LAL 01-01 (January 4th 2001).

\bibitem{3r} S. Hagelin, Nucl. Phys. {\bf B193}, 123 (1981).
 
  \bibitem{4r} A. Le Yaouanc, L. Oliver, O. P\`ene and J.-C. Raynal, 
Phys. Lett. {\bf B316}, 567 (1993).

\bibitem{5r} A. M. Shifman and M. Voloshin, Sov. J. Nucl. Phys. {\bf 
47}, 511 (1988).
 

\bibitem{6r} V. Mor\'enas, A. Le Yaouanc, L. Oliver, O. P\`ene and 
J.-C. Raynal, Phys. Rev. {\bf D58}, 114019 (1998).
 
\bibitem{10r} B. Bakamjian and L. H. Thomas, Phys. Rev. {\bf
92}, 1300 (1953)~; A. Le Yaouanc, L. Oliver, O. P\`ene and J.-C. 
Raynal, Phys. Lett. {\bf B365}, 319 (1996).

\bibitem{14r} S. Godfrey and N. Isgur, Phys. Rev. {\bf D32}, 189 (1985).

\end{thebibliography}
\end{document}